\documentclass[12pt,twoside]{article}
\usepackage{epsfig}
\newcommand{\VolumeHeader}{}
\newcommand{\VolumeSerial}{}

\newcommand{\ActivityName}{ {\normalsize {\it 
Summer School on  Astroparticle Physics and Cosmology
}}}
\newcommand{\ActivityDate}{ {\normalsize {\it
ICTP, Trieste, 17 June - 5 July 2002 
}}}

\newcommand{\ndt}{\noindent}

\newcommand{\ov}{\overline}

\newcommand{\be}{\begin{equation}}
\newcommand{\ee}{\end{equation}}
\newcommand{\bea}{\begin{eqnarray}}
\newcommand{\eea}{\end{eqnarray}}
\newcommand{\LectureHeader}{MM Searches}

\begin{document}
\pagestyle{myheadings}
\markboth{\LectureHeader}{\VolumeHeader}
\markright{\VolumeHeader}

%%%%%%%%%%%%%%%%%%%%%%%%%%%%%%%%%%%%%%%%%%%%%%%%%%%%%%%%%%%%%%%%%%%%%%%%%%%
%%%            Title page starts here                                     %
%%%%%%%%%%%%%%%%%%%%%%%%%%%%%%%%%%%%%%%%%%%%%%%%%%%%%%%%%%%%%%%%%%%%%%%%%%%

\begin{titlepage}

%%% YOUR CHANGES BELOW THIS LINE

\title{Magnetic Monopole Searches} 

\author{G. Giacomelli$^\dagger$ and L. Patrizii$^\dagger$ \thanks{giacomelli@bo.infn.it; patrizii@bo.infn.it}
\\[1cm]
{\normalsize
{\it $^\dagger$ Dipartimento di Fisica dell'Universit\`a di Bologna}}
 \\ 
{\normalsize
{\it and INFN,
 Sezione di Bologna }}
 \\ 
{\normalsize 
{\it Viale C. Berti Pichat 6/2, I-40127, Bologna, Italy}}
\\[10cm]
%%% FOR FURTHER AUTHORS SEE WHAT IT IS WRITTEN IN THE ABSTRACT 
%%% DO NOT CHANGE THE FOLLOWING LINES
{\normalsize {\it Lecture given at the: }}
\\
\ActivityName 
\\
\ActivityDate 
\\[1cm]
{\small \VolumeSerial} 
}
\date{}
\maketitle
\thispagestyle{empty}
\end{titlepage}

\baselineskip=14pt
\newpage
\thispagestyle{empty}

%%%%%%%%%%%%%%%%%%%%%%%%%%%%%%%%%%%%%%%%%%%%%%%%%%%%%%%%%%%%%%%%%%%%%%%%%%%
%%%            Abstract page starts here                                  %
%%%%%%%%%%%%%%%%%%%%%%%%%%%%%%%%%%%%%%%%%%%%%%%%%%%%%%%%%%%%%%%%%%%%%%%%%%%

\begin{abstract}

%%% YOUR CHANGES BELOW THIS LINE
In this lecture notes will be discussed  the status of the  
searches (i) for classical Dirac Magnetic Monopoles (MMs) at accelerators, 
(ii) for GUT superheavy MMs in the penetrating cosmic radiation 
and  (iii) for Intermediate Mass MMs in the cosmic radiation 
underground, underwater and at high altitude. An outlook and a 
discussion on future searches follows.
\end{abstract}

\vspace{6cm}



%%%%%%%%%%%%%%%%%%%%%%%%%%%%%%%%%%%%%%%%%%%%%%%%%%%%%%%%%%%%%%%%%%%%%%%%%%%
%%%       Automatic TOC and your Text starts here                         %
%%%%%%%%%%%%%%%%%%%%%%%%%%%%%%%%%%%%%%%%%%%%%%%%%%%%%%%%%%%%%%%%%%%%%%%%%%%

\newpage
\thispagestyle{empty}
\tableofcontents

\newpage
\setcounter{page}{1}

\section{Introduction}

In 1931 Dirac introduced the magnetic monopole in order to explain the 
quantization of the electric charge, which
 follows from the existence of at least one free magnetic 
charge \cite{dirac}. He established the basic relationship between 
the  
elementary electric
charge $e$ and the basic magnetic charge $g$
\begin{equation}
	eg=n\hbar c/2
\end{equation}
where $n$ is an integer, $n=1,2,..$. The magnetic charge is 
$g = n g_{D}$;\ 
 $g_{D}=\hbar c/2e = 68.5 e$ is called the unit Dirac charge. The existence of magnetic charges and of magnetic currents would symmetrize in form the Maxwell's equations, but the symmetry would not be perfect since $e \neq g$. But the couplings could perhaps be energy dependent and they could merge in a single common value at very high energies \cite{derujula}.\par
   %The basic
%properties of a MM follow from the Dirac relation. In particular 
%because of the large magnetic charge, a MM acquires a large 
%energy in a magnetic field, and a fast MM ($\beta>10^{-2}$) behaves 
%as an equivalent electric charge $(ze)_{eq}=g_D\beta$,  with $\beta=v/c$. 
There was no prediction
 for the MM mass; from 1931  searches for  
``classical Dirac monopoles'' were carried out at every new 
accelerator using mainly  relatively simple set—-ups, and recently 
also large collider detectors [3-9]. Searches at the Fermilab collider seem to exclude MMs with masses  up to 850 GeV. Experiments at the LEP2 collider exclude masses below  102 GeV \cite{opal}.
\\\indent
Electric charge is naturally quantized in GUT  gauge theories of the 
basic interactions; such theories imply the existence of MMs, with 
calculable properties. The MMs appear in the Early Universe at 
the phase transition 
corresponding to the spontaneous breaking of the unified group 
into subgroups,
one of which is U(1) \cite{thooft}. The  MM     
mass is related to the mass  of  the carriers X, Y of the
unified interaction, $ m_{M}\ge m_{X}/G$, 
where G is the dimensionless unified coupling constant at energies E 
$\simeq m_{X}$. 
In GUTs with  
$m_{X}\simeq 10^{14}-10^{15}$ GeV and $G\simeq 0.025$, $m_{M}> 
10^{16}-10^{17}$ GeV. 
This is an enormous
mass: MMs cannot be produced at any man--made accelerator, 
existing or conceivable. They could only be produced in the first 
instants of 
our  Universe and can be searched  for in the 
penetrating Cosmic Radiation (CR). 
\\\indent
 Larger MM masses are expected
 if gravity is brought into the unification 
 picture, and in some  SuperSymmetric models.
 \\\indent
Intermediate mass monopoles (IMMs) may have been produced in later
 phase transitions in the Early Universe, in which a semisimple 
gauge group
yields a U(1) group \cite{lazaride}. IMMs with masses $10^{5} \div 10^{12}$ GeV 
may be accelerated to relativistic velocities in the galactic 
magnetic field, and in several astrophysical sites. It has been speculated that very energetic IMMs could yield the highest energy cosmic rays \cite{bhatta}.
\\\indent
The lowest mass MM should be stable, since magnetic charge is 
conserved like electric charge. Therefore, the MMs produced in 
the Early Universe should still exist as cosmic relics, whose 
kinetic energy has been affected first by the expansion of the 
Universe and then by their travel through galactic and 
intergalactic magnetic fields. 
\\\indent
GUT poles in the CR should have low velocities and 
relatively large energy losses; they are best searched for 
underground in the penetrating cosmic radiation. IMMs could be 
relativistic and may be searched for at high altitude laboratories, 
in the downgoing CR and, if very energetic, also in the upgoing 
CR.
\\
\\
%\vspace{3mm}
 \indent
  In this lecture we shall review the present experimental situation on MM searches with emphasis on classical Dirac monopoles and on Intermediate Mass MMs. Recently there has been renewed interest in the search for relatively low--mass Dirac MMs. In fact there seem to be no a priori reasons that Dirac MMs (or dyons) might not exist \cite{anti-d0}.\par
\section{Main properties of magnetic monopoles}
The main properties of MMs  are obtained from the Dirac relation (1), and are 
 summarized  here. We recall that the Dirac relation may be easily obtained 
semiclassically by considering the system of one monopole and one 
electron, and quantizing the radial component of the total angular 
momentum. \par 
\noindent - {\it Magnetic charge.} 
If $n$~=1 and if the basic electric charge is that of the 
electron, then  
the basic magnetic charge is 
$ g_D =\hbar c/ 2e=137e/2=3.29\times 10^{-8} \ cgs=\ 68.5 e$. The magnetic charge should be larger if  $n>1$ and also if the basic electric charge is $e/3$.

\noindent - {\it Coupling constant.} 
In analogy with the fine structure constant, $\alpha 
=e^{2}/\hbar c\simeq 
1/137$, the 
 dimensionless magnetic coupling constant is 
$ \alpha_g=g^{2}_{D}/ \hbar c \simeq 34.25$; notice that it is very large, much larger than 1, and thus perturbative methods cannot be used.
\par
\ndt - {\it Energy W acquired in a magnetic field  B}:~  
$  W = ng_{D} B\ell = n \ 20.5$ keV/G~cm.
In a coherent galactic--length   
  ($\ell\simeq 1$ kpc, and $B\simeq 3~\mu$G), the energy gained  
by a monopole is
 $ W \simeq 1.8\times 10^{11}$ GeV.
 Classical poles and IMMs in the cosmic radiation could 
be 
accelerated to relativistic velocities. 
  Instead GUT poles have large masses and are expected 
to have relatively low velocities, $10^{-4}<\beta<10^{-1}$. 
\par                 
\noindent- {\it Trapping of MMs in ferromagnetic materials.}  
MMs may be
trapped in  ferromagnetic materials by an image force, 
which  could reach the value of $\simeq 10$ eV/\AA.
\par
\noindent- Electrically
charged monopoles (dyons) may arise as quantum--mechanical 
excitations or as M--p, M-nucleus composites.\par
\noindent- There is no real prediction of the mass of classical Dirac
MMs. One may have a rough estimate assuming that the classical 
monopole radius is equal to the classical electron radius:
$r_M= \frac{g^{2}}{m_Mc^{2}}= r_e=\frac{e^{2}}{m_ec^{2}}$, from which $m_M=\frac{g^{2}m_e}{e^{2}} \simeq n \ 4700\  m_e \simeq n \ 2.4\  GeV/c^{2}$. Thus the mass should be relatively large and even larger   
  if the basic  charge is $e/3$ and if $n>1$.
\vspace{0.2cm} 
\par
Also the interactions of MMs with matter  are connected with the 
electromagnetic properties of MMs and thus are consequences of 
the Dirac relation. It is also important to know whether 
the quantity and quality of
energy lost by a MM in a particle detector is adequate for 
its 
detection. 
 The interaction of the MM magnetic charge with nuclear magnetic 
dipoles 
 could lead 
to the formation of M--nucleus bound systems. 
 This may affect the energy loss
in matter and the cross--section for MM catalysis of proton 
decay.
 A monopole--proton bound state may be produced via  radiative 
capture,  
$ M+p\to    (M+p)_{bound} + \gamma$.
Monopole--nucleus bound states may exist for nuclei with a 
large gyromagnetic ratio.
 \par
 \noindent- {\it Energy losses of fast poles.} 
A fast MM with magnetic charge $g_D$ and velocity $v=\beta c$ 
behaves like an equivalent electric charge 
$(ze)_{eq}=g_D\beta$; the energy losses of fast monopoles are 
thus very large.\par
%\ndt - {\it Energy losses of fast poles.} A fast MM 
%moving  with velocity $v>10^{-2}c$ behaves like an equivalent 
%electric
%charge $(Ze)^{2}_{eq}=g^{2}\beta^{2}$. \par
\noindent - {\it Energy losses of slow monopoles} ($10^{-
4}<\beta<10^{-2}$).
 For slow particles it is important to distinguish the energy 
lost   
in ionization or  excitation of atoms and molecules of the medium 
(``electronic'' energy
loss) from that lost to yield kinetic energy to recoiling atoms 
or 
nuclei  
(``atomic'' or ``nuclear'' energy loss). Electronic energy loss 
predominates for 
 electrically or magnetically charged particles 
 with $\beta> 10^{-3}$. 
 The dE/dx of MMs  with $10^{-4}<\beta<10^{-3}$
is mainly due to excitations of 
atoms. In an ionization detector using noble gases there would be, for $10^{-4}<\beta<10^{-3}$, an additional energy loss due to atomic energy level mixing (Drell effect, see Section 3).
 \par
\noindent - {\it Energy losses at very low velocities.} 
MMs with   $v<10^{-4}c$ cannot excite atoms;
they can only lose energy in elastic collisions with atoms or 
with nuclei.
 The energy is released to the 
medium in the form 
of elastic vibrations and/or infra--red radiation \cite{derkaoui1}.\par
\begin{figure}[h]
	\begin{center}
		\includegraphics[width=0.8\textwidth]{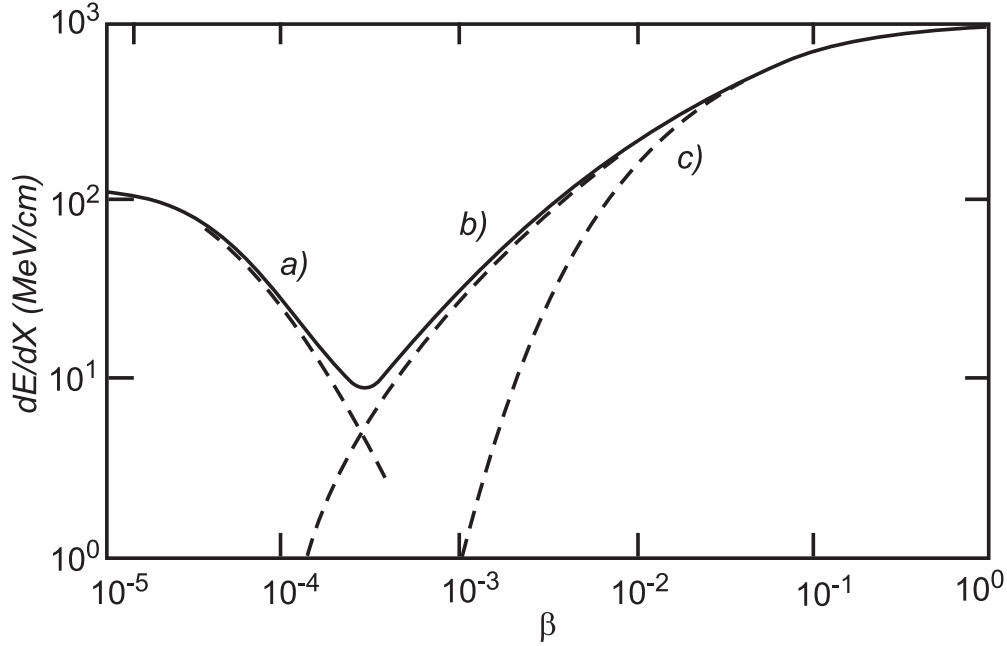}
	\end{center}
	\caption{The energy losses, in MeV/cm, of $g=g_D$ MMs in
liquid hydrogen as a function of ${ \beta}$. Curve a) corresponds
to elastic monopole--hydrogen atom scattering; curve b) 
corresponds
to interactions with level crossings; curve c) describes
the ionization energy loss.}
	\label{fig:perdita-di-energia}
\end{figure}

Fig.\ \ref{fig:perdita-di-energia} shows a sketch of the   energy losses in liquid hydrogen  
of
a $g=g_D$ MM vs its $\beta$ \cite{gg+lp}.\par

\noindent - {\it Energy losses in superconductors.}
 If a pole
passes through a superconducting ring, there will be a magnetic 
flux change of $\phi_B=2\pi\hbar c/e$, 
 yielding $dE/dx\simeq 42$~MeV/cm, $\beta-$independent (see Section 3).\par
\noindent - {\it  Energy losses of MMs in celestial bodies.}
 For  $\beta$ $<10^{-4}$  the main energy losses in 
the Earth are due to : i) pole--atom elastic scattering, 
ii) eddy current losses, 
iii) nuclear stopping power.
Poles may be stopped by celestial bodies if they have\\
\noindent Moon: $\beta\leq 5\times {10^{-5}}$,\quad 
Earth: $\beta \leq 10^{-4}$,\quad   
Jupiter: $\beta \leq 3\times {10^{-4}}$,\quad Sun: $\beta \leq 
10^{-3}.$\par  

\section{Monopole detectors}
\par
Monopole detectors are based  on the properties of MMs determined from Dirac's 
relation. \par
\noindent - {\it  Superconducting induction devices.}
This method of
detection  is based only on the long--range 
electromagnetic interaction between the magnetic charge and the 
macroscopic 
quantum state
of a superconducting ring. 
A moving MM 
induces in the ring  an 
electromotive force and a current ($\Delta i$).
 For a  coil with N turns and inductance 
{\it L},  $ \Delta i=4\pi N ng_D/L=2\Delta i_o$, 
where $\Delta i_o$ is the current change corresponding to a 
change of one unit 
of the flux quantum of superconductivity.
 A superconducting induction detector,
 consisting of a detection coil coupled to a SQUID 
(Superconducting Quantum
Interferometer Device), should be sensitive to MMs of any 
velocity \cite{gg1}. 

\noindent - {\it Scintillation counters.}
 Many searches have  been performed 
using excitation loss techniques. 
The light
yield from a MM traversing a  scintillator has a threshold at 
$\beta \sim 10^{-4}$, above which the light signal 
is  large compared
to that of a minimum ionizing particle. 
For $10^{-3}< \beta < 10^{-1}$ there is a saturation effect. For 
$\beta >0.1$ the light yield 
 increases because of the production of many delta rays \cite{derkaoui2,macro1}. 

\noindent - {\it Gaseous detectors.} 
Gaseous detectors of various types have been used. 
MACRO used limited streamer tubes equipped
with readouts for the wires and pickup strips, for two--dimensional 
localization \cite{macro2}. 
The gas   was 73\% helium and 27\% n--pentane. This
 allows  exploitation 
of the Drell \cite{drell} and Penning effects: a magnetic monopole leaves the 
helium atoms in a metastable excited state (He*) with an excited 
energy of  
$\simeq 20$ eV. The ionization potential of n--pentane is about 10 
eV;  the 
Penning effect 
converts the excited energy of the He*  into ionization of the n--pentane 
molecule \cite{gg1,gg}. \par

\noindent - {\it Nuclear track detectors.} Nuclear track detectors  (NTD) can record the passage of heavily ionizing particles like magnetic monopoles \cite{oujda}.
The formation of an etchable track in a nuclear track detector is related to the Restricted
Energy Loss
(REL), which is the fraction of the total energy loss which remains localized
in a cylindrical region with about 10 nm diameter
around the particle trajectory.
Both the electronic  and the nuclear energy losses contribute to REL. In Ref. \cite{cr39} it was shown that both are effective in producing etchable tracks in the CR39 nuclear track detector. The CR39 has a threshold at $z/\beta \simeq5$; it is the most sensitive NTD and it  allows to search for magnetic monopoles
with one unit Dirac charge (g=$g_D$) for $\beta$ around $10^{-4}$
and for  $\beta >10^{-3}$, the whole  $\beta$-range of
$4 \times 10^{-5}<\beta< 1$ for MMs with $g \geq 2 g_D$ \cite{derkaoui2}. The Lexan and Makrofol polycarbonates have a threshold at $z/\beta \sim 50$; thus they are sensitive only to relativistic MMs.

\section{Searches for ``classical Dirac monopoles''}
We shall consider ``classical" Dirac monopoles those MMs which have relatively low
masses and could possibly be produced at accelerators.\par
\begin{figure}[h]
	\begin{center}
		\includegraphics[width=0.8\textwidth]{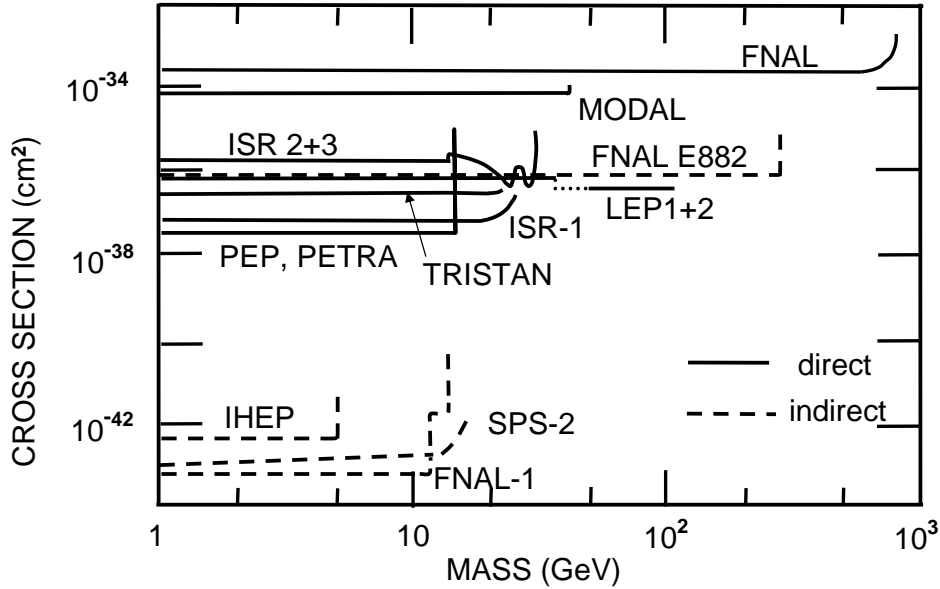}
	\end{center}
	\caption{Cross section upper limits vs MM mass obtained 
from direct accelerator searches  (solid lines) and indirect 
searches (dashed lines).}
%\vspace{3mm}
	\label{fig:mmclass2}
\end{figure}

\noindent - {\it Accelerator searches.} 
If MMs could be produced at high--energy accelerators, they would 
be 
 re\-la\-ti\-vi\-stic and  would ionize heavily. They would thus be 
easily discriminated from minimum ionizing particles.
 Examples of direct searches are  scintillation counter searches 
and 
the  experiments performed 
with nuclear track detectors for which data taking is integrated 
 over periods of
 months. Experiments at the Fermilab $\overline p p$ collider 
 established cross section
upper limits of $\sim 2\times 10^{-34}$~cm$^2$ for MMs with 
masses up to 
850 GeV \cite{gg,bertani}.
 Searches at $e^{+}e^{-}$  colliders excluded 
masses up to 45 GeV 
\cite{gg} and later also the 45-102 GeV mass range (the cross section upper limits are at $\sigma \sim 5\times 10^{-37}$~cm$^2$) \cite{opal}, 
Fig.\ \ref{fig:mmclass2}. Recently several large purpose detectors at high energy colliders have used some of their subdetectors (mainly the tracking subdetectors) to search for classical Dirac monopoles.\par
\begin{figure}[h]
	\begin{center}
		\includegraphics[width=0.76\textwidth]{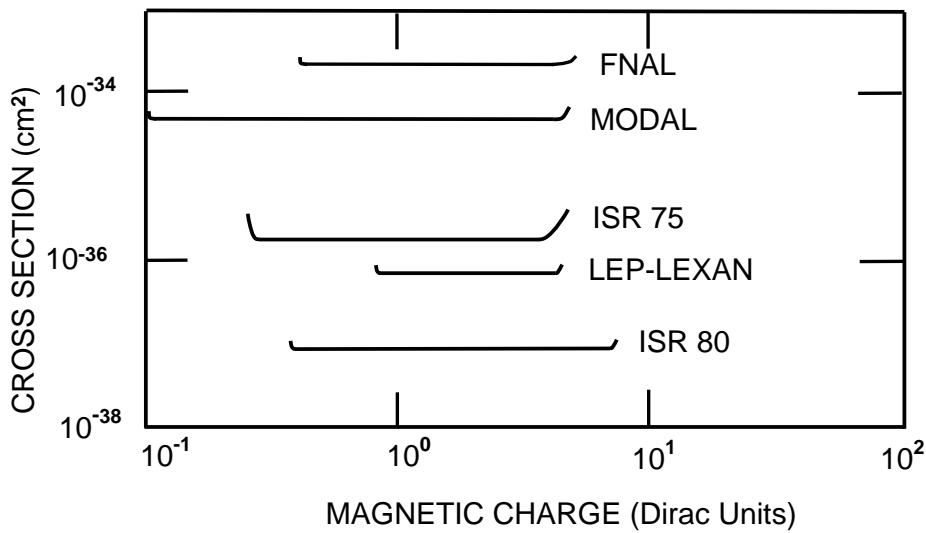}
	\end{center}
	\caption{Upper limits (95\% C.L.) for classical--monopole production for some direct experiments plotted versus magnetic charge.}
	\label{fig:fig3}
\end{figure}

Fig.\ \ref{fig:fig3} summarizes the direct limits as a function of the monopole magnetic charge $g =n g_{D}/q$ (the value of 1 corresponds to the electric charge of the electron and $n=1$); if the basic electric charge is that of a quark with $q=1/3$, then the magnetic charge would be 3 times larger.\par
An example of  an indirect search is an experiment at the CERN 
SPS: the 450 GeV
protons interacted  in  targets made
of  ferromagnetic tungsten powder. 
Later on the targets were placed in
front of a pulsed solenoid with a field 
$B\sim 200$ kG,  large enough to extract  and
 accelerate the MMs, to be  detected in nuclear 
emulsions and in  CR39  sheets \cite{gg1}. A more recent indirect experiment was performed at the $\overline p p$ Tevatron collider at Fermilab, assuming that the produced MMs could stop, be trapped and bound in matter surrounding the D0 collision region \cite{kalbfleish}. Beryllium and Aluminium samples of the materials having dimensions of $\leq$ 7.5 cm diameter and 7.5 cm long, were repeatedly passed through the 10 cm diameter warm bore centered on and perpendicular to two superconducting coils. The induced charge (current) in the superconducting coil could be measured by DC SQUIDs. Monopole mass limits $m>285$ GeV were established for $g=g_D$ poles. It is difficult to establish the validity of several hypotheses which have to be used in order to interpret these negative results.

\vspace{9mm}

\ndt - {\it Multi--$\gamma$ events.} 
Five peculiar photon shower events, found in nuclear plates 
exposed to 
high--altitude cosmic rays, were characterized by 
an  energetic narrow cone of tens of photons, without any 
incident charged 
particle \cite{multigamma}. The total energy in the photons was of the order of 
$10^{11}$ GeV. The
small 
radial spread of photons  suggested a c.m. $\gamma=(1-\beta^{2})^{-1/2}>10^3$. 
The energies of the photons in the overall c.m. 
system were  small, too low to have $\pi^o$ decays as 
their source. 
One possible explanation of these events could be the following: 
a 
high--energy $\gamma$--ray, with energy  $>10^{12}$ eV, produced 
in the
plate a pole--antipole pair, which then suffered bremsstrahlung 
and annihilation 
producing the final multi--$\gamma$ events. \par
\begin{figure}[h]
	\begin{center}
		\includegraphics[width=0.670\textwidth]{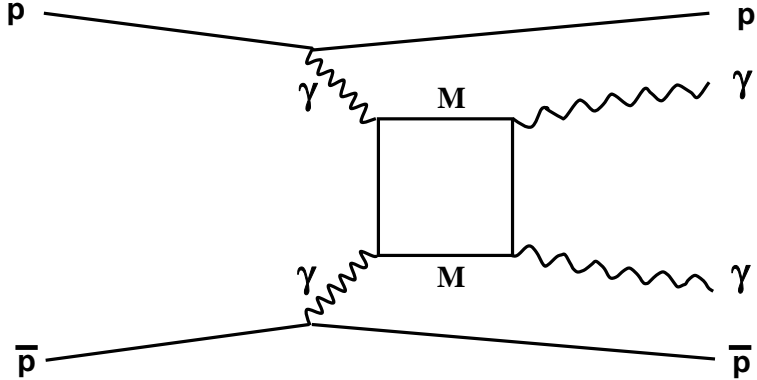}
	\end{center}
	\caption{Feynman diagram for $\gamma\gamma$ producion via a  virtual monopole loop in $p \ov p$ collisions at the Tevatron. The $\gamma\gamma \rightarrow \gamma\gamma$ process cross sections at energies  below the magnetic monopole production threshold could be enhanced due to the strong coupling of the virtual MMs to photons [6].}
	\label{fig:feynman2}
\end{figure}

  Searches for multi-$\gamma$ events were  performed in $pp$ collisions at the ISR 
at $\sqrt{s}=53$ GeV \cite{gg}, in $\ov p p$ collisions  at  the 1.8 TeV collider at Fermilab  and in $e^{+}e^{-}$ collisions at LEP. The ISR experiment placed a  cross--section  upper--limit of 
$\sim 10^{-37}$ cm$^2$.  At Fermilab the D0 experiment searched for pairs of photons with high transverse energies; virtual heavy pointlike Dirac MMs could rescatter pairs of nearly real photons into the final state via a box monopole diagram as shown in Fig.\ \ref{fig:feynman2}. They set a 95\% C.L. lower limit of 870 GeV for spin 1/2 Dirac monopoles \cite{d0}.  At LEP the L3 collaboration searched for anomalous $Z\rightarrow \gamma\gamma\gamma$ events; they observed no significant deviation from QED predictions, setting a 95\% C.L. lower mass limit of 510 GeV \cite{l3}. Many authors studied the effects from virtual monopole loops \cite{derujula,ginzburg}.
The authors of Ref.~\cite{anti-d0} criticized the underlying theory and believe that no significant limit can be obtained from present experiments based on virtual monopole processes.

\noindent - {\it Searches in bulk matter.} 
 Classical MMs could be produced  by cosmic 
rays and 
 could stop at the surface of
the Earth, where they could be trapped in ferromagnetic 
 materials.
 A search 
for MMs in bulk matter used a total of 331 kg of 
material, including meteorites, schists, ferromanganese 
nodules, iron ore and other materials. The detector was a 
superconducting 
induction coil with a SQUID.
The material was passed at constant velocity through the 
magnet bore. The passage of a MM 
trapped in a sample would cause a jump in the current in the 
superconducting coil.
 From the 
absence of candidates the authors conclude that the 
monopole/nucleon ratio in 
the samples was $<1.2\times 10^{-29}$ at 90\% C.L. \cite{gg1}. \par
The searches for classical MMs performed at accelerators are not relevant to 
the question of the existence of very massive poles.
Ruzicka and Zrelov  summarized  all
searches for classical monopoles performed before 1980 
\cite{ruzicka}. A more recent bibliography, until the end of 
1999, is given in Ref.~\cite{biblio}. Possible effects arising from low 
mass MMs have been reported \cite{oscuro}.

\par

\section{Supermassive GUT monopoles}
As  already stated, GUT theories of the electroweak and strong 
interations predict the existence of superheavy magnetic monopoles 
produced in the Early Universe (EU) as topological point defects when a
GUT gauge group breaks into separate groups, one of which is 
U(1). Assuming that the GUT group is SU(5) (in reality it is excluded by 
proton decay experiments) one should have the following transitions in the EU:
\begin{equation}
%\scriptsize
%\begin{displaymath}
    \begin{array}{ccccc}
        {} & 10^{15}\ GeV & {} & 10^{2}\ GeV & {} \\
        SU(5) & \longrightarrow & SU(3)_{C}\times \left[ SU(2)_{L}\times U(1)_{Y}\right] & \longrightarrow & SU(3)_{C}\times U(1)_{EM} \\
       {} & \small10^{-35}s & {} & \small10^{-9}s & {}
    \end{array}
%\end{displaymath}
%\normalsize
\end{equation}
MMs would be generated as topological point defects in the GUT phase transition, about
one monopole for each causal domain. In the 
standard cosmology this leads to too many monopoles: the present monopole 
density would be $\rho_{M}\sim 5 \times 10^{-18}$  g/cm$^{3}$, while the critical density is $\rho_{c}\sim 8 \times 10^{-29}$  g/cm$^{3}$ (the 
monopole problem!). Inflation would defer the GUT phase 
transition, after extreme supercooling; in its simplest version 
the number of generated MMs would be very small. However the flux depends critically on several parameters, like $m_{M}$, the reheating temperature, etc. If the reheating temperature is large enough one would have MMs produced in high energy collisions, like $e^{+}e^{-}\rightarrow M\ov M$. \par
\begin{figure}[h]
	\begin{center}
		\includegraphics[width=0.74\textwidth]{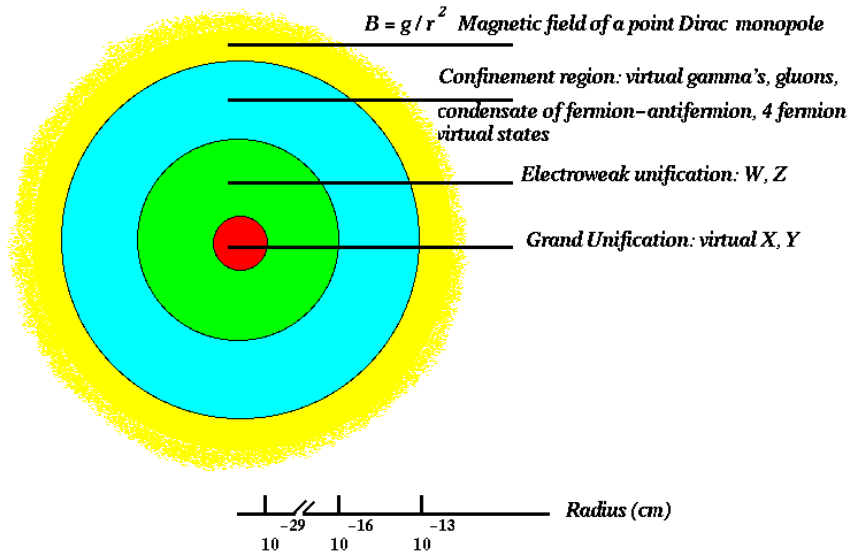}
	\end{center}
	\caption{Figure~5: Structure of a GUT monopole. The various 
regions correspond
to: (i) Grand Unification ($r \sim 10^{-29}$ cm; inside this core 
one finds
virtual $X$ and $Y$ particles); (ii) electroweak unification
($r \sim 10^{-16}$ cm; inside one finds virtual 
$W^{\pm}$
and $Z^0$); (iii)
confinement region ($r \sim 10^{-13}$ cm; inside one finds 
virtual
$\gamma$, gluons and a condensate of fermion-antifermion pairs 
and possibly 4-fermion
virtual states); (iv) for $r>$ few fm one has the field of a 
point
magnetic charge.}
	\label{fig:gut}
\end{figure}

 
Fig.\ \ref{fig:gut} shows the possible structure of a GUT magnetic monopole, with a very small core, an electroweak region,  a confinement region, a fermion--antifermion condensate region (which may contain 4--fermion baryon--number--violating 
terms);
for $r\geq  3$ fm a MM behaves as a  point particle  
 which generates a field $B=g/r^{2}$ \cite{picture}.\par

\begin{figure}[h]
	\begin{center}
		\includegraphics[width=0.82\textwidth]{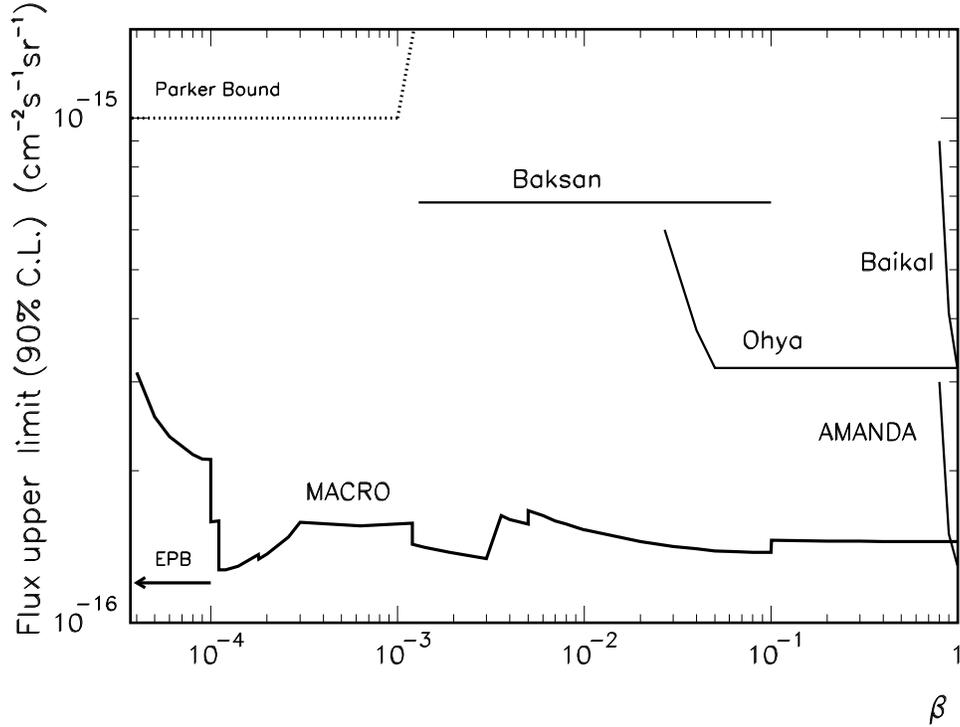}
	\end{center}
	\caption{The 90\% C.L. global MACRO direct upper limits vs $\beta$ for GUT  $g=g_D$ monopoles in the penetrating CR, compared with limits from other experiments [28-31].}
	\label{fig:global2}
\end{figure}

A flux of cosmic GUT supermassive magnetic monopoles may reach 
the Earth
and may have done so for the whole life of the Earth. The 
velocity
spectrum of these MMs  could be in the range $4 \times 10^{-5} 
<\beta <0.1$,
with possible peaks corresponding to the escape velocities from 
the Earth,
the Sun and the Galaxy.
Searches for such MMs in the  cosmic radiation 
have been performed with superconducting induction
devices, whose combined limit is at the level of
$2 \times 10^{-14}~$cm$^{-2}$~s$^{-1}$~sr$^{-1}$, independent of 
$\beta$ \cite{gg}.
Several direct searches were performed above ground and 
underground [18, 28-31].
 %\cite{gg,ohya}. 
The most complete search was 
performed by the MACRO detector, using  three different types of 
subdetectors (liquid scintillators, limited streamer tubes and nuclear track detectors) and
with an acceptance of about 10,000 m$^2$sr for an isotropic flux.
 No
monopoles have been detected;  the  90\% C.L. flux
limits are shown  in
Fig.\ \ref{fig:global2} vs $\beta$  for $g=g_D$ MMs \cite{mm_macro}: the limits are  at the level of 
$1.4\times 10^{-16}$~cm$^{-2}$~s$^{-1}$~sr$^{-1}$ for $\beta > 4 \times 10^{-5}$. The figure shows also the limits from the Ohya \cite{ohya}, Baksan, Baikal, and AMANDA experiments \cite{baksan}. Previous limits are at levels larger than  10$^{-15}$~cm$^{-2}$~s$^{-1}$~sr$^{-1}$ \cite{mm_macro}. \par

\begin{figure}[h]
	\begin{center}
		\includegraphics[width=0.8\textwidth]{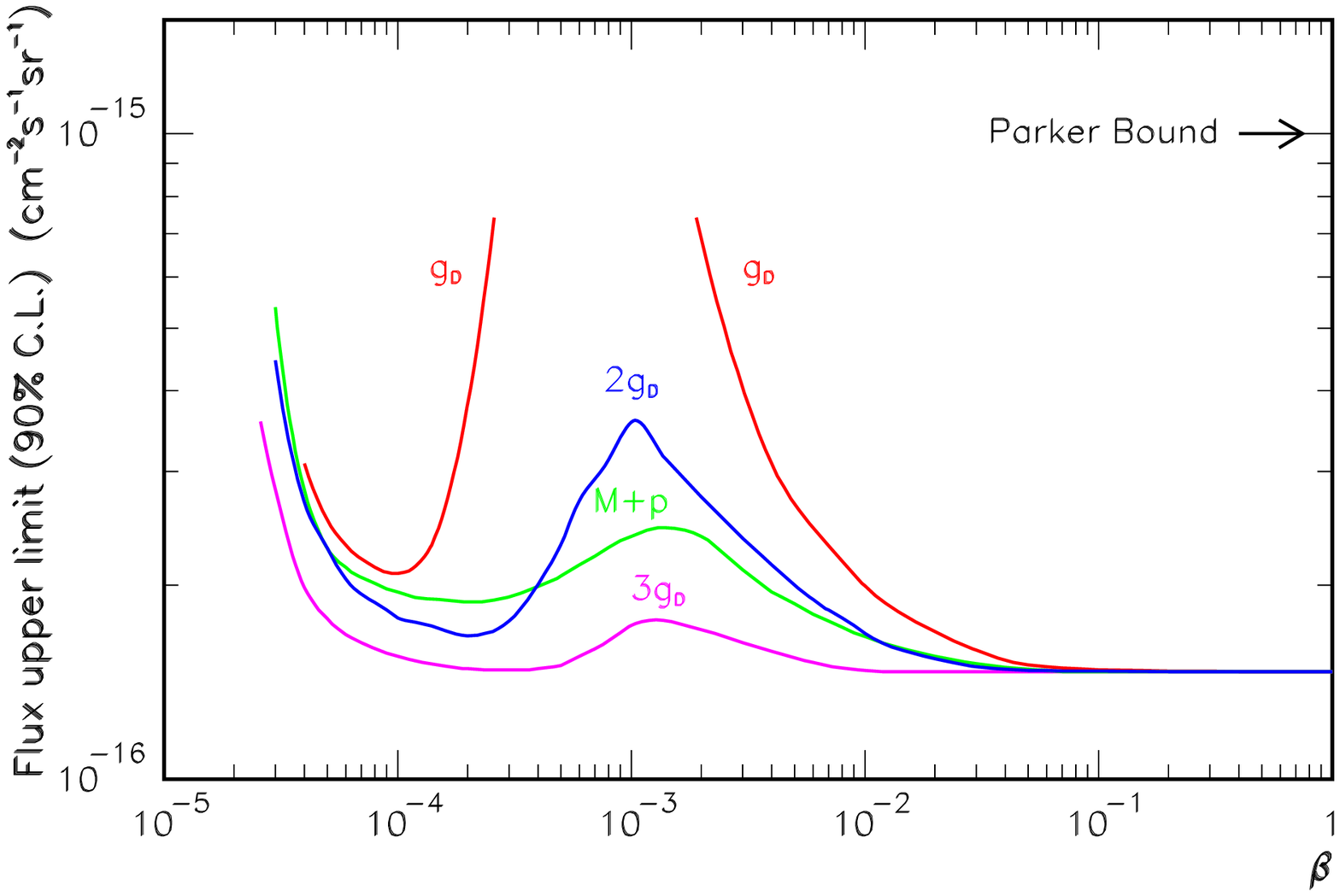}
	\end{center}
	\caption{Upper limits (90\% C.L.) for an isotropic flux of MMs in the cosmic radiation, obtained with the CR39 subdetector of MACRO, for poles with magnetic charges \(g=g_D,\; 2g_D, \;3g_D\) and for M+p composites.}
	\label{fig:cr39final}
\end{figure}

Fig.\ \ref{fig:cr39final} shows the 90\% C.L. flux upper limits  obtained with the MACRO CR39 nuclear track detector for MMs with different magnetic 
charges, \(g=g_{D}\), \( 2g_{D}\), \(3g_{D} \) and for M + p composites \cite{gg02}.

\par
\begin{figure}[h]
	\begin{center}
		\includegraphics[width=0.65\textwidth]{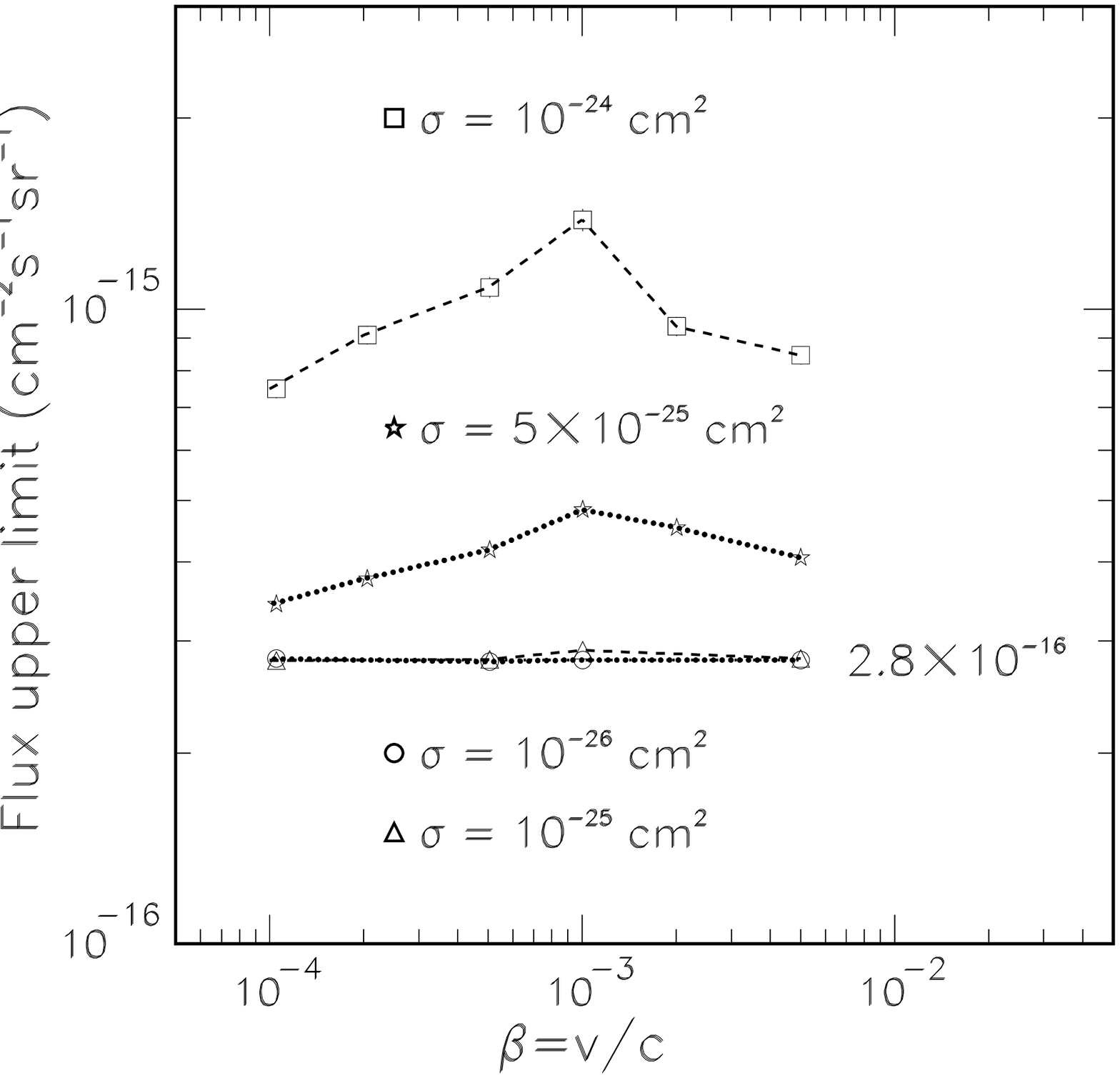}
	\end{center}
	\caption{The MACRO 90\% C.L. upper limits for a MM flux as a function of the MM velocity for various catalysis cross sections [31]. The limit from the Baikal underwater detector is $\Phi \le 6 \times 10^{-17} $cm$^{-2}$s$^{-1}$sr$^{-1}$ for $\beta \simeq 10^{-5}$ [30, 33].}
	\label{fig:catalisi}
\end{figure}

 
%\vspace{2mm}
The interaction of the GUT monopole core with a nucleon can lead to a reaction in which the nucleon decays (monopole catalysis of nucleon decay), f. e. \( M + p \rightarrow M + e^+ + \pi^0\). The cross 
section for this process is of the order of magnitude of the core size, \( \sigma \sim 10^{-56}$ cm$^2\), 
practically negligible. But the catalysis process could proceed via the Rubakov-Callan mechanism with a cross section of the order  of the  strong interaction cross section \cite{rubakov}. MACRO developed a dedicated analysis procedure aiming to detect nucleon decays induced by the passage of a GUT monopole in their streamer tube system. The flux 
upper limit results of this search as a function of the MM velocity and of the catalysis cross section are shown in Fig.\ \ref{fig:catalisi} \cite{catalisi}.
 Previous limits are at levels larger than $10^{-15}$~cm$^{-2}$~s$^{-1}$~sr$^{-1}$ \cite{catalisi}, with the exception of the Baikal limit which is $ 6 \times 10^{-17}$~cm$^{-2}$~s$^{-1}$~sr$^{-1}$
 for $\beta \simeq 10^{-5}$ \cite{baksan}.

%\vspace{2mm}
Some indirect searches  used ancient  mica, which has a high 
z threshold. 
 The mica experiment 
scenario assumes that a bare monopole passing through the Earth 
captures an 
aluminium nucleus and drags it through subterranean mica causing 
a trail of 
lattice defects. As long as the mica is not reheated, the damage 
trail will
survive. The  mica pieces analyzed are small 
($13.5$ and $18$ cm$^2$), but should have been recording tracks 
since they cooled,  
about $4\div9\times 10^8$ years ago. The flux
  upper--limits  are at the level of 
$10^{-17} ~\mbox{cm}^{-2}~ \mbox{s}^{-1} $sr$^{-1}$ for $10^{-
4}<\beta<10^{-3}$
\cite{price}.
There are many reasons why these indirect experiments 
 might not be 
sensitive. For example, if MMs have a positive electric 
charge  or have protons attached, then Coulomb repulsion could 
prevent capture of heavy nuclei.\par

\section{Cosmological and astrophysical bounds}
Rough, order of magnitude upper limits for a GUT monopole flux in 
the cosmic radiation 
were obtained on the basis of cosmological 
and astrophysical considerations. Here we shall quote only some of these limits.\par

\ndt - {\it Limit from the mass density of the universe.} 
 This
 bound is obtained requiring that the present MM mass density 
be 
smaller than the critical density $\rho_c$ of the universe. 
 For $m_M\simeq 10^{17}$ GeV one has the  
 limit:
 $ F={n_Mc\over 4\pi}\beta<3\times  
10^{-12}h^2_0\beta~(\mbox{cm}^{-2}\mbox{s}^{-1} \mbox{sr}^{-1}).$
 It is valid for poles uniformly distributed in the universe. If 
poles are 
clustered in galaxies the  limit could be much larger.

\ndt - {\it Limit from the galactic magnetic field. The Parker 
limit.}
 The $\sim 3\  \mu$G magnetic field 
in our Galaxy is stretched in the  direction of the spiral
 arms; 
it is probably due to the non--uniform rotation of the Galaxy. 
This 
mechanism generates a field with a time--scale approximately 
equal to 
the rotation period of the Galaxy $(\tau\sim 10^8$ yr). Since MMs  
are accelerated in magnetic fields, they  gain energy, which is 
taken away
from the stored magnetic energy. An upper bound for the MM
flux is  
obtained by requiring that the kinetic energy gained per unit 
time by 
 MMs  be less than or  
equal to  the magnetic energy generated by the dynamo 
effect. This yields the so--called Parker limit: $F<10^{-
15}~\mbox{cm}^{-2}~\mbox{s}^{-1}$ sr$^{-1}$ \cite{parker}. 
The original limit was re--examined to take into
 account  the 
almost  
chaotic nature of the galactic magnetic field, with domain 
lengths of about 
$\ell\sim 1$ kpc; the limit becomes  mass 
dependent \cite{parker}. 
More recently an extended Parker bound was obtained by 
considering the survival of 
an early seed field \cite{adams}. The result was
$ F\leq 1.2 \times 10^{-16}(m_M/10^{17}GeV)~\mbox{cm}^{-
2}~\mbox{s}^{-1}~
\mbox{sr}^{-1}.$
\par
\ndt - {\it Limit from the intergalactic magnetic field.}
Assuming the existence in the local group of galaxies of an 
intergalactic field $B_{IG}\sim 3\times 10^{-8}~G$ with a 
regeneration time 
$\tau_{IG}\sim 10^9~y$ and  applying the same reasoning discussed 
above, a 
  more 
stringent  bound is obtained;
 the
 limit is less reliable because the intergalactic field is less 
known.
\par
\ndt - {\it Limits  from peculiar A4 stars and from pulsars.}
Peculiar A4 stars have their magnetic fields 
$(B\sim 10^3~G)$ in the direction opposite to that expected from 
their 
rotation. 
 A MM with $\beta\leq 10^{-3}$ would  stop in A4 stars;
 thus the number of MMs in the 
star would increase with time (neglecting $M\ov M$ annihilations 
inside the 
star). The poles could be accelerated in the magnetic field, 
which would 
therefore decrease with increasing time. Repeating the Parker 
argument, 
 one may obtain 
%$ F<3\times 10^{-20}\mbox{cm}^{-2}~\mbox{s}^{-1}$ sr$^{-1}$.  
 strong limits, but it is not clear how good are all the 
assumptions made. 
With similar considerations applied to the superconducting core 
of neutron stars, the 
field survival of a pulsar gives an upper limit of the monopole 
flux in the 
neighbourhood of the pulsar. 
The limit would be particulary stringent for pulsar PSR 1937+214 \cite{gg1,gg+lp}.

\section{Intermediate mass magnetic monopoles}
IMMs would appear as topological point defects at a later time in the Early Universe; in this case the GUT group would not yield a U(1) group at the end of the GUT phase transition, it would appear a later new phase transition, as for instance  in the   following sequence 

\begin{equation}
%\scriptsize
%\begin{displaymath}
    \begin{array}{ccccc}
        {} & 10^{15}\ GeV  & {}& 10^{9}\ GeV & \\
        SO(10) & \longrightarrow & SU(4)\times SU(2)\times SU(2) & \longrightarrow & SU(3)\times SU(2)\times U(1) \\
        {} & \small10^{-35}s & {} & \small10^{-23}s & {}
    \end{array}
%\end{displaymath}
%\normalsize
\end{equation}

\noindent which would lead to MMs with masses of the order of  $10^{10}$ GeV; these monopoles  survive inflation, are stable, ``doubly charged'' (n=2 in Eq.1) and do not catalyze nucleon decay \cite{lazaride}.
The structure of an IMM would be similar to that of a 
GUT 
monopole, but the core would be much larger (since R $\sim$ 1/$m_M$) 
and the outer 
cloud would not contain 4--fermion baryon--number--violating 
terms. \par
Relativistic magnetic monopoles with intermediate masses, 
$10^5<m_M<10^{12}$ 
GeV, could
be present in the cosmic radiation. IMMs could  be  accelerated to large  values of $\gamma$ in one coherent domain of the galactic magnetic field. Thus one would have to look for $\beta \ge 0.1$ fast, heavily ionizing MMs.\par
Detectors underground, underwater and under ice  would mainly have a sensitivity  for poles coming from above.
 Detectors at the Earth 
surface 
could  detect MMs
coming from above if they have masses larger than $10^5-
10^6$ GeV \cite{derkaoui1}; lower
mass MMs may be searched for with detectors located at high 
mountain
altitudes, or  in balloons and in satellites. \par
\begin{figure}[ht]
	\begin{center}
		\includegraphics[width=0.80\textwidth]{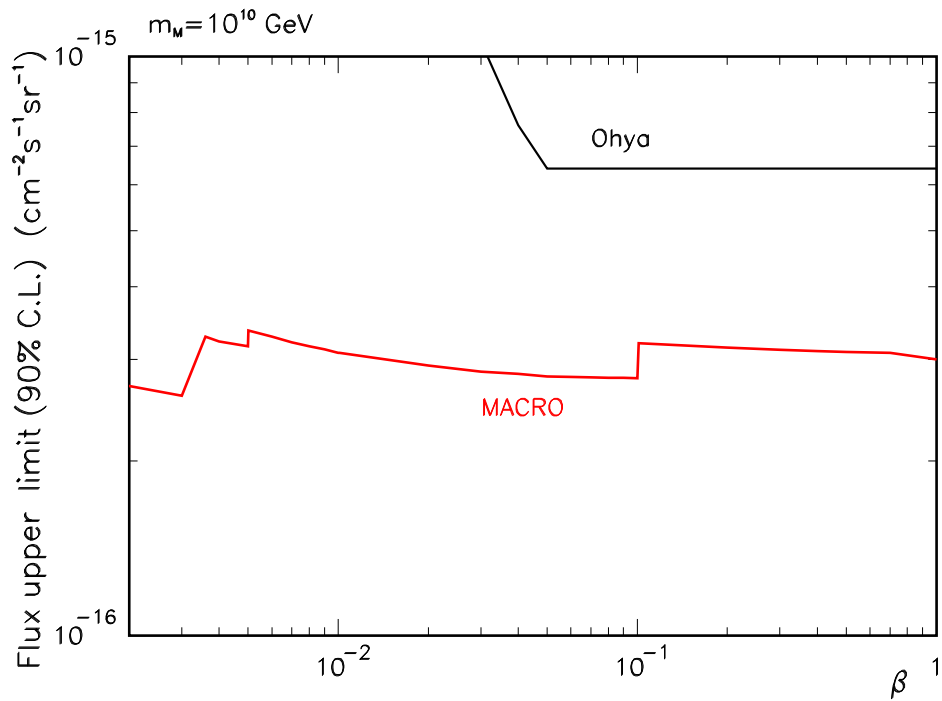}
	\end{center}
	\caption{Experimental 90\% C.L. upper limits for a flux of IMMs with mass $m_M=10^{10}$ GeV  plotted versus $\beta$.}
	\label{fig:imm1}
\end{figure}

 Few 
experimental
results are available \cite{nakamura}. Fig.\ \ref{fig:imm1} shows the present situation on flux upper limits for intermediate mass MMs.  The Cherenkov neutrino telescopes under ice and underwater are sensitive to fast and very fast ($\gamma >>1$) MMs mainly coming from above.

The SLIM experiment is searching for fast IMMs
with nuclear track detectors at the Chacaltaya
high altitude lab (5230 m above sea level) \cite{slim}. It is sensitive to MMs with $4 \times 10^{-5}<\beta <3\times 10^{-4}$ and $\beta> 2 \times 10^{-3}$ if $g=g_D$, the whole range $4 \times 10^{-5}<\beta <1$ if $g=2g_D$.
 Nuclear track detectors are sensitive to these poles and are also sensitive to slow moving nuclearites (strangelets, strange quark matter).

\section{Conclusions. Outlook}
\vspace{-0.5cm}$    $\par
Direct and indirect accelerator searches for classical Dirac monopoles 
have placed 95 \% C.L. mass 
limits at the level of $m_M > 850$ GeV with cross section upper values as shown in Fig.\ \ref{fig:mmclass2}. Future improvements could 
come from
experiments at the LHC \cite{moedal}.
\\
\indent Many searches have been performed for superheavy GUT monopoles in the penetrating cosmic radiation. The 90 \% C.L. flux limits are at the level of 
$ \Phi \le 1.4 \times 10^{-16} $~cm$^{-2}$~s$^{-1}$~sr$^{-1}$ for $\beta \ge 
4 \times 10^{-5}$.
It would be difficult to do much better since one would require refined
detectors of
considerably larger areas. Or one has to devise completely new 
techniques.\\
\indent
Present limits on Intermediate Mass Monopoles with high $\beta$ are relatively poor. 
Experiments at high altitudes and at neutrino telescopes should
 improve  the situation. In particular stringent limits may be obtained by large neutrino telescopes for IMMs with $\beta > 0.5$ coming from above. \par
 As a byproduct of GUT MM searches  MACRO  obtained stringent limits on nuclearites in the CR \cite{gg02}. Future experiments at neutrino telescopes and at high altitude should  perform searches for smaller mass nuclearites.\par

\section*{Acknowledgments}

We would like to acknowledge the cooperation of many colleagues, in particular S. Cecchini, F. Cei, M. Cozzi, I. De Mitri, M. Giorgini, G. Mandrioli, M. Ouchrif, V. Popa, P. Serra, M. Spurio, and others.
%%%%%%%%%%%%%%%%%%%%%%%%%%%%%%%%%%%%%%%%%%%%%%%%%%%%%%%%%%%%%%%%%%%%%%%%%%%

%%%%%%%%%%%%%%%%%%%%%%%%%%%%%%%%%%%%%%%%%%%%%%%%%%%%%%%%%%%%%%%%%%%%%%%%%%%
%%%           References starts here                                      %
%%%%%%%%%%%%%%%%%%%%%%%%%%%%%%%%%%%%%%%%%%%%%%%%%%%%%%%%%%%%%%%%%%%%%%%%%%%

\end{document}